\documentclass[conference]{IEEEtran}
\pdfoutput=1
\usepackage[table]{xcolor}
\usepackage{cite}
\usepackage{amsmath,amssymb,amsfonts}
\usepackage{algorithmic}
\usepackage{graphicx}
\usepackage{textcomp}
\usepackage{xcolor}
\usepackage{verbatim}
\usepackage{microtype}

\def\BibTeX{{\rm B\kern-.05em{\sc i\kern-.025em b}\kern-.08em
    T\kern-.1667em\lower.7ex\hbox{E}\kern-.125em X}}
\begin{document}

\makeatletter
\newcommand{\linebreakand}{%
  \end{@IEEEauthorhalign}
  \hfill\mbox{}\par
  \mbox{}\hfill\begin{@IEEEauthorhalign}
}
\makeatother

\title{Particle Mesh Ewald for Molecular Dynamics \\ 
in OpenCL on an FPGA Cluster}

\author{\IEEEauthorblockN{Lawrence C. Stewart}
\IEEEauthorblockA{\textit{Silicon Therapeutics} \\
451 D St, Suite 205\\
Boston, MA, USA \\
larry.stewart@silicontx.com}\\
\and
\IEEEauthorblockN{Carlo Pascoe}
\IEEEauthorblockA{\textit{Silicon Therapeutics}\\ 
451 D St, Suite 205\\
Boston, MA, USA \\
carlo.pascoe@silicontx.com}\\
\and
\IEEEauthorblockN{Emery Davis}
\IEEEauthorblockA{\textit{Silicon Therapeutics}\\ 
451 D St, Suite 205\\
Boston, MA, USA \\
emery.davis@silicontx.com}\\

\linebreakand 
\IEEEauthorblockN{Brian W. Sherman}
\IEEEauthorblockA{\textit{Silicon Therapeutics} \\
451 D St, Suite 205\\
Boston, MA, USA \\
woody@silicontx.com}\\
\and
\IEEEauthorblockN{Martin Herbordt}
\IEEEauthorblockA{\textit{College of Engineering} \\
\textit{Boston University}\\
Boston MA, USA\\
herbordt@bu.edu}\\
\and
\IEEEauthorblockN{Vipin Sachdeva}
\IEEEauthorblockA{\textit{Silicon Therapeutics} \\
451 D St, Suite 205 \\
Boston MA, USA\\
vipin@silicontx.com}
}

\setlength{\intextsep}{6pt plus 2pt}   

  \maketitle

\vspace{-10pt}

   \begin{abstract}

Molecular Dynamics (MD) simulations play a central role in physics-driven drug discovery. MD applications often use the Particle Mesh Ewald (PME) algorithm to accelerate electrostatic force computations, but efficient parallelization has proven difficult due to the high communication requirements of distributed 3D FFTs. In this paper, we present the design and implementation of a scalable PME algorithm that runs on a cluster of Intel Stratix 10 FPGAs and can handle FFT sizes appropriate to address real-world drug discovery projects (grids up to {\boldmath $128^3$}). To our knowledge, this is the first work to fully integrate all aspects of the PME algorithm (charge spreading, 3D FFT/IFFT, and force interpolation) within a distributed FPGA framework. The design is fully implemented with OpenCL for flexibility and ease of development and uses 100 Gbps links for direct FPGA-to-FPGA communications without the need for host interaction. We present experimental data up to 4 FPGAs (e.g., 206 microseconds per timestep for a 65536 atom simulation and {\boldmath $64^3$} 3D FFT), outperforming GPUs for smaller FFT sizes. Additionally, we discuss design scalability on clusters with differing topologies up to 64 FPGAs (with expected performance far greater than all known GPU implementations) and integration with other hardware components to form a complete molecular dynamics application. We predict best-case performance of 6.6 microseconds per timestep on 64 FPGAs.
\end{abstract}
\begin{IEEEkeywords}
FPGA, Molecular Dynamics, HPC, Reconfigurable Computing, FFT
\end{IEEEkeywords}
  \section{Introduction} \label{introduction}

Molecular dynamics (MD) simulation engines such as AMBER~\cite{amberoverview} and OpenMM~\cite{eastman2017openmm} provide high performance implementations for CPU and GPU, and provide a flexible framework in which new computational technologies can be assessed. FPGA implementations have also been explored for many years \cite{Azizi04,Kindratenko06,Scrofano06a,Alam07} including a recent study showing promising single-FPGA performance \cite{yang2019fully}. 

MD plays a critical role in computational chemistry in general and in drug discovery in particular. There, long timescales and small problem sizes lead to tremendous challenges in strong scaling, especially in the electrostatics computation. This has led to the creation of ASIC-based solutions \cite{ohmura2014mdgrape,anton1,shaw2014anton}; their limitations, however, due to cost and availability, make COTS alternatives essential. Of these, only FPGA clusters have shown potential past a small number of nodes \cite{sheng14,lawandediss}. These preliminary studies, however, were based only on the parallelization of the 3D FFT and not full electrostatics, much less complete system integration.

The electrostatic force computation often uses Ewald summation to split the work into short-range and long-range terms. Methods for the long-range term include k-space summation\cite{halver2020kokkos}, $\mu$-series\cite{predescu2019}, and use of Fourier Transforms to solve Poisson's Equation\cite{essmann1995smooth}. The FFT methods reduce computation from ${O(N^2)}$ to ${O(N log N)}$, but require global communication and strong scaling of 3D FFTs in the size range from $32^3$ to $128^3$ as well as a complex mapping function \cite{Sanaullah16}.

In this paper we describe the architecture, implementation, and evaluation of a distributed 3D FFT-based Smooth Particle Mesh Ewald electrostatic force computation. Our implementation outperforms GPUs in the problem size ranges of interest for drug discovery, and is scalable to multiple pipelines on multiple boards. It is implemented entirely in OpenCL. There are a number of contributions.
\begin{itemize}
    \item This is the first parallel complete FPGA electrostatics.
    \item This work provides a still-rare case study of a production HPC application successfully implemented in OpenCL and distributed across a parallel cluster of tightly coupled FPGAs.
    \item To our knowledge, this effort is the first to obtain strong scaling for MD problems by using off the shelf hardware. The system is integrated into a complete MD application and results validated against OpenMM over millions of timesteps.
\end{itemize}

The potential impact is as follows. Many desirable drug targets in cancer, auto-immune, neurodegenerative, and infectious diseases are currently considered undruggable due lack of binding predictions (binding free energies, conformational changes) \cite{Stockwell11} that could be provided with long timescale MD \cite{Klepeis09}. The current work will enable these timescales to be achieved an order of magnitude faster.





The outline of this paper is as follows: Section~\ref{sec:relatedwork} discusses background and related work on MD and FFT targeting contemporary architectures including CPUs, GPUs, and ASICs. Section~\ref{sec:architecture} details the system architecture of our long-range pipeline and 3D FFT, and Section~\ref{sec:implementation} follows up with the implementation details. 
Section~\ref{sec:evaluation} summarizes the results of our work, along with performance comparison to other hardware. Finally, Section~\ref{sec:conclusion} concludes the paper and describes our plans for future work. 

  \section{Background and Related Work}


Molecular Dynamics models the behavior of atoms and molecules by individually calculating the various forces that act on them. Forces that apply to bonded atoms include bond torsions and tensions. Forces that apply to non-bonded atoms include short-range forces that include both van der Waals and electrostatics, and long-range forces, which are mainly electrostatic. Short-range forces are managed by pairwise computations. For long-range forces, applications instead use multipole approximations\cite{greengard1997fast} or Ewald summations. Our focus is on an Ewald variation known as Smooth Particle Mesh Ewald~(SPME)~\cite{spme}. SPME calculates a charge distribution on a grid, then uses Fourier Transforms and a Green's function to calculate a potential field. Potential gradients then are used to calculate forces. 

\label{sec:relatedwork}

Molecular dynamics simulations have proven to be a valuable tool in drug discovery for understanding protein motion. Open-source GPU accelerated molecular dynamics applications such as GROMACS~\cite{berendsen1995gromacs}, NAMD~\cite{phillips2005scalable}, OpenMM~\cite{eastman2017openmm}, and CP2K\cite{kuhne2020cp2k} allow many practitioners to use MD simulations as a regular tool. 
To our knowledge, the only study showing strong scaling on multiple GPUs for a 100,000 atom system is with the recently redeveloped GROMACS package~\cite{nvidiablog,v100perf}, which does not distribute the FFT.

Several efforts to develop, at great expense, custom ASICs for small molecule simulations have been undertaken. The earliest initiative is the MDGRAPE~\cite{ohmura2014mdgrape} series of supercomputers.
Another well known ASIC initiative is the Anton series~\cite{anton1} developed by D. E. Shaw Research. Anton 1 was released in 2007, with performance for a 23,000 atom system close to 17 microseconds/day. Anton 2, released in 2014, increased this performance five-fold to 85 microseconds/day~\cite{shaw2014anton}. 

The most challenging part of scaling molecular dynamics simulations is the electrostatic forces computation, of which FFT is often a major component.
Anton 1 could solve FFT problems of size $32^3$ in 3.7 microseconds, and $64^3$ in 13.3 microseconds on 512 nodes. Anton 2 did not use FFT in its simulations, instead relying on a different decomposition called the  $\mu$-series~\cite{shaw2014anton,predescu2019}. 

Efforts to get parts of molecular dynamics simulations running on FPGAs have been explored over the past few years~\cite{khan2013fpga-accelerated,chiu10}. More recently, the increase in FPGA resources such as logic elements, DSPs, BRAM, etc., have allowed full MD simulations to run on a single FPGA~\cite{yang2019fully}.

A great strength of FPGAs is the I/O transceivers, which are capable of providing a great deal of bandwidth with very low latency~\cite{IntelStratix10}. Some clusters with highly interconnected FPGAs are the Novo-G\# built at the University of Florida in a 3D torus interconnect~\cite{George16} and the first version of the Microsoft Catapult~\cite{Putnam14}. More recently, University of Paderborn has developed Noctua~\cite{plessl2018bringing} and Tsukuba University has deployed Cygnus, a hybrid GPU-FPGA system~\cite{kobayashi2018opencl}. 
FPGA communications can also now be programmed using OpenCL, providing both high-performance and a productive development environment for distributed applications. 
Prior work on 3D FFTs on single FPGAs includes~\cite{Sasaki05,Yu11,Humphries14,kuhne2020cp2k,fft3d-fpga} while work on multiple FPGAs includes~\cite{sheng14,lawandediss}. Design of FFT for MD simulations is presented in~\cite{sheng17}. The earliest 2D floating point FFT on multiple FPGAs of which we are aware is \cite{lee2008multi}. 


\section{System Architecture} \label{systemarchitecture}
\label{sec:architecture}

Figure~\ref{fig:MDOverview} shows the OpenCL portions of our modified OpenMM application and the role played by 3D FFT. Our goal is to run multiple timesteps of the full MD application on a network of FPGAs without any additional host communication beyond initialization and result collection.
The focus of this paper is the long range portion of the system.

\begin{figure}[tb]
  \centering
  \includegraphics[width=\linewidth]{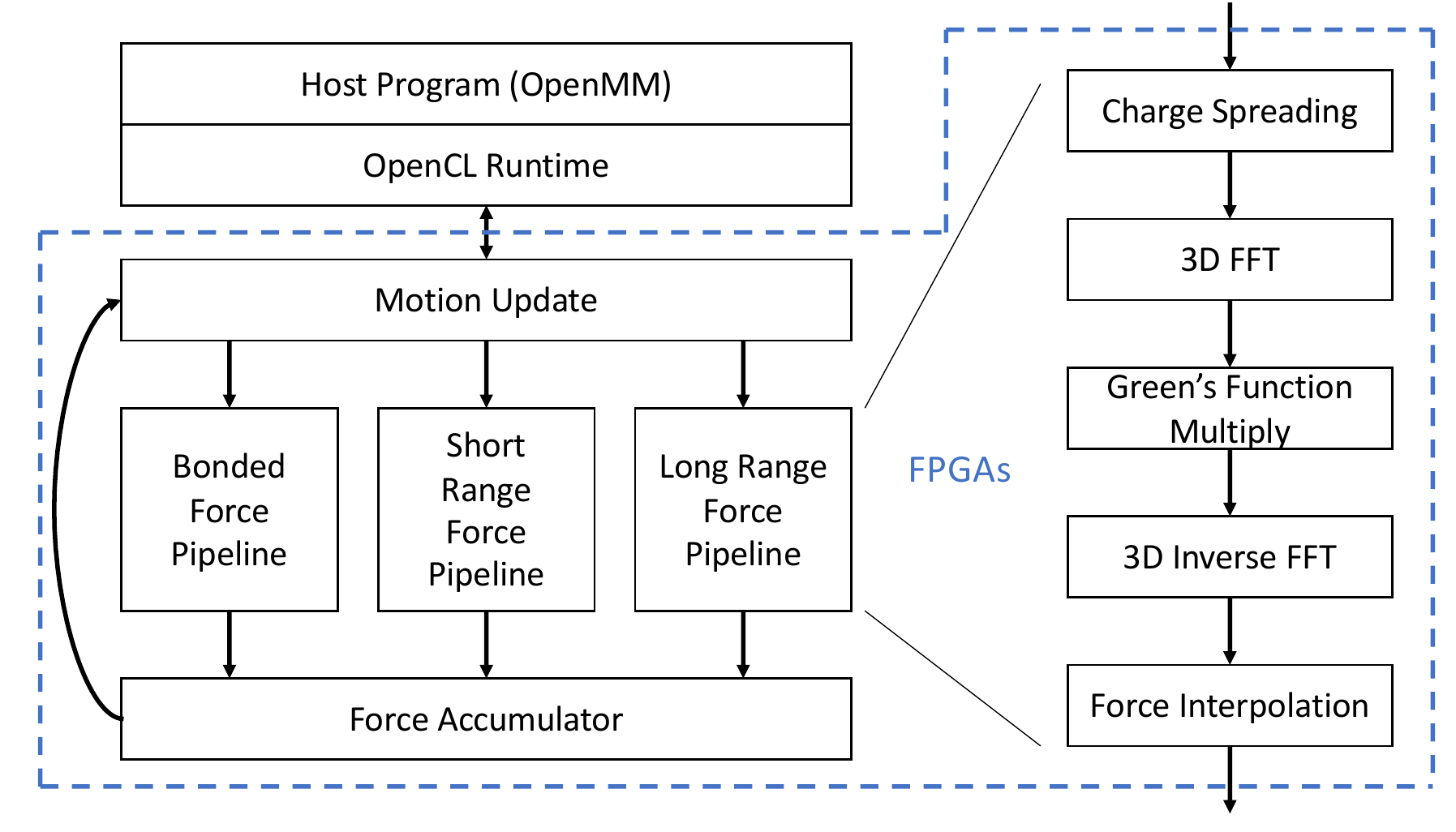}
  \vspace*{-0.2truein}
  \caption{Molecular dynamics application overview.}
  \label{fig:MDOverview}
  \vspace*{-0.1truein}
\end{figure}

\subsection{Long-Range Pipeline}

The architecture of the long-range force pipeline is shown in Figure~\ref{fig:LRPipe}. Each timestep, the LR pipeline accepts atom positions and charges as input, and delivers long-range electrostatic force updates per atom back to the force accumulator and motion update portions of the system. 

\begin{figure}[t]
  \centering
  \includegraphics[width=\linewidth]{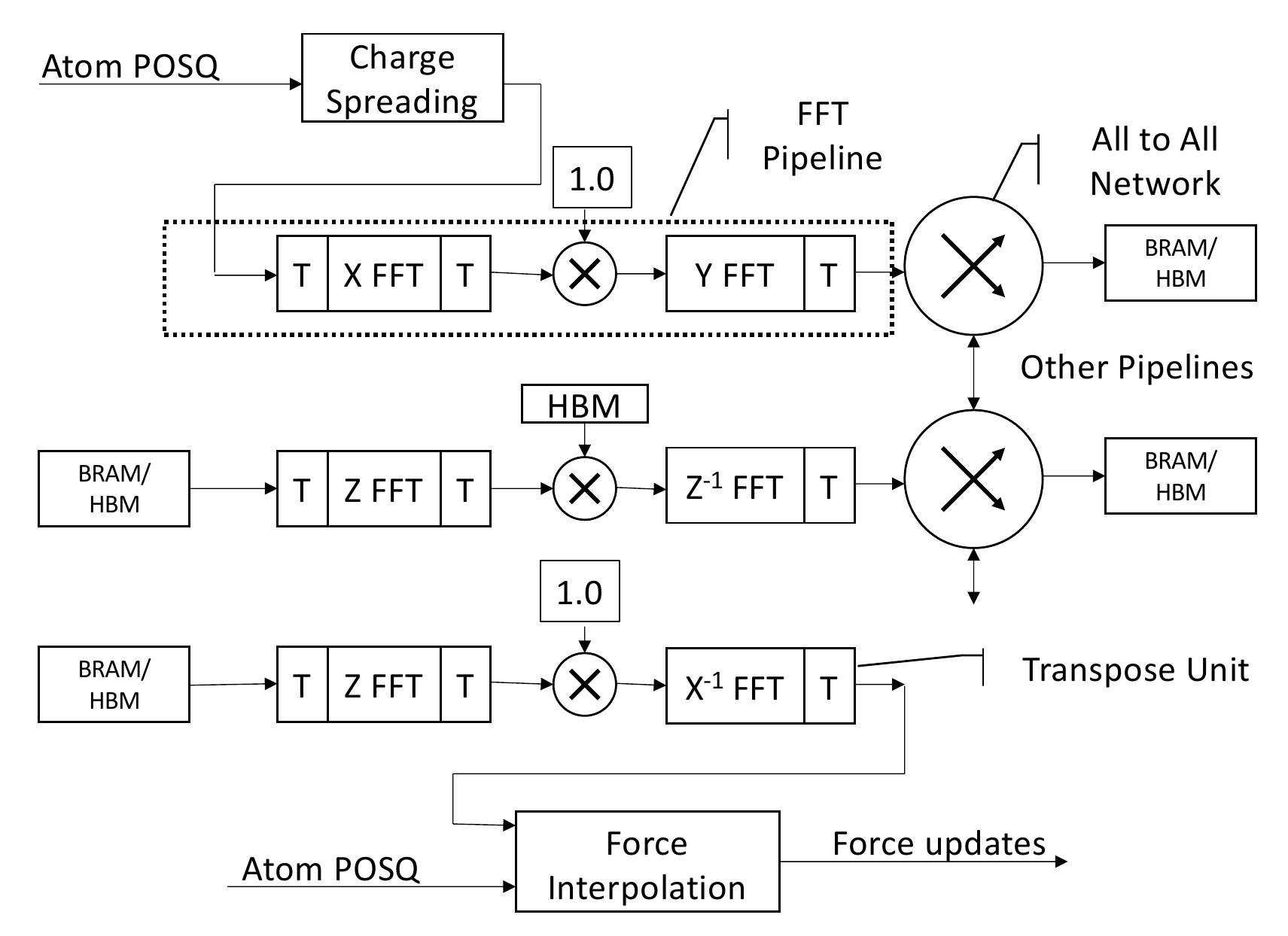}
  \vspace*{-0.2truein}
  \caption{Logical view of long-range force pipeline.}
  \label{fig:LRPipe}
  \vspace*{-10pt}
\end{figure}



\subsubsection{Charge Spreading}

Charge Spreading accepts atom positions and charges and constructs the FFT input volume. Any particular atom has a fractional position between grid points. The charge attached to the atom is spread to grid points in the 4x4x4 cell surrounding the atom's position by using cardinal B-splines as interpolating functions.  We use 64-way parallel hardware to achieve a single pipeline charge spreading throughput of one atom per clock cycle. The computation is further parallelized by multiple pipelines and multiple FPGAs.  Because atom volumes of influence overlap, we use additional hardware to dynamically reorder atoms to avoid pipeline hazards. Atoms which affect multiple pipelines are processed by each affected pipe.

\subsubsection{FFT} 

The 3D FFT is described in more detail in section~\ref{systemimplementation}.  Our implementation divides the input volume into slabs in the Z dimension.  Each slab runs independently to compute X and Y transforms, then an all-to-all network exchanges data -- called corner-turning -- so that the slabs subdivide the Y dimension.  Parallel pipelines then compute Z direction transforms, multiply by Green's function data, and compute $Z^{-1}$ transforms.  A second corner turn feeds $Y^{-1}$ and $X^{-1}$ transforms.


\subsubsection{Force Interpolation}

The force interpolation unit shares the 64-way parallel arithmetic and BRAM design of the charge spreading hardware. It accepts data from the output of the inverse FFT, and atom position and charge data. 3D cardinal B-splines and their derivatives are used to calculate X, Y, and Z forces for each atom based on gradients of the electrostatic potential field calculated by the FFT. 64-way combining trees calculate force updates which are fed back to the force accumulation block of the main MD application. We plan to share hardware between charge spreading and force interpolation, but this is not yet done.

\subsection{Scaling 3D FFT}

The three dimensional FFT of an XYZ volume can be computed as a sequence of one dimensional transforms~\cite{Tolimieri:fft,frigo2005design}. 
There are alternative parallel 3D FFT formulations such as 2D decomposition \cite{pekurovsky2012p3dfft} and the generalized vector radix decomposition\cite{harris1977vector} but for up to 64 nodes and $128^3$ our initial focus is on the simple decomposition.

3D FFT can be scaled by using pipelined parallel hardware, and then by using multiple parallel compute units, provided that the necessary operands can be routed to the compute units.

In 2013, Garrido et al. showed how to build pipelined parallel FFT hardware using a feedforward architecture that is well suited to FPGA implementations~\cite{garrido2011pipelined}. 
A single precision complex 8-wide FFT unit uses 4\% of a Stratix 10 device depending on transform size and consumes and delivers about 19 GB/sec of data.


This hardware can complete 1D FFTs in the number of clock cycles it takes to read the data. Using a nominal 300 MHz design speed, 
Table~\ref{table:fftusec} shows the time in microseconds to complete $N^2$ 1D FFTs for various size transforms, given different numbers of 8-wide vector compute units. To complete a full forward and inverse 3D FFT will take six times as much work but can be both pipelined and parallelized.

In order to distribute such a system over a network of FPGAs, it is necessary to balance communications and computation performance and it is also necessary to choose points in the solution spaces for FFT and for All to All in which the bandwidths match.

\begin{table*}[t!]
\caption{Time for FFT sizes vs number of units, at 300 MHz ($\mu$s)}
\vspace{-6pt}
\centering
{%
\begin{tabular}{|c|r|r|c|c|c|c|c|c|c|c|}
\hline
XYZ&Data Points&Data Bits&1&2&4&8&16&32&64&128 \\
\hline
32x32x32 & 32,768 & 2,097,152 & 13.7&6.8&3.4&1.7&0.9&0.4&0.2&0.1 \\
64x64x64 & 262,144 & 16,777,216 &109.2&54.6&27.3&13.7&6.8&3.4&1.7&0.9 \\
64x64x128 & 524,288 & 33,554,432 &218.5&109.2&54.6&27.3&13.7&6.8&3.4&1.7 \\
96x96x96 & 884,736 & 56,623,104 &368.6&184.3&92.2&46.1&23.0&11.5&5.8&2.9 \\
128x128x128 & 2,097,152 & 134,217,728 &873.8&436.9&218.5&109.2 (1)&54.6 (2)&27.3 (3)&13.7(4)&6.8 (5) \\
\hline
\end{tabular}}
{\smallskip \\
Data within () reference similar entries in table~\ref{table:a2anetwork}}
\label{table:fftusec}
\vspace{-10pt}
\end{table*}

\begin{table*}[t!]
\caption{Network timing for All to All ($\mu$s)}
\vspace{-6pt}
\centering
{%

\begin{tabular}{|c|c|c|c|c|c|c|c|c|c|}
\hline
Nodes & Topology & Hopcount & Links & 32x32x32 & 64x64x64 & 64x64x128 & 96x96x96 & 128x128x128 \\
\hline
2 & PTOP & 1 & 4 & 1.7 & 13.4 & 26.9 & 45.4 & 107.5 (1) \\
4 & PtoP & 1 & 3 & 1.7 & 13.4 & 26.9 & 45.4 & 107.5 (1) \\
8 & 2D Torus & 1.5 & 4 & 1.1 & 8.8 & 17.6 & 29.8 & 70.6 \\
8 & Hypercube & 1.5 & 3 & 1.5 & 11.8 & 23.5 & 39.7 & 94.1 \\
8 & Hypercube++ & 1.25 & 4 & 0.9 & 7.4 & 14.7 & 24.8 & 58.8 (2)\\
8 & 3D Torus & 1.5 & 3 & 1.5 & 11.8 & 23.5 & 39.7 & 47.1 \\
8 & Switched & 1 & 4 & 0.7 & 5.9 & 11.8 & 19.8 & 47.1 \\
16 & 2D Torus & 2 & 4 & 0.8 & 6.3 & 12.6 & 21.3 & 50.4 \\
16 & 3D Torus & 2 & 6 & 0.5 & 4.2 & 8.4 & 14.2 & 33.6 (3) \\
16 & Hypercube & 2 & 4 & 0.8 & 6.3 & 12.6 & 21.3 & 50.4 \\
16 & Switched & 1 & 4 & 0.4 & 3.2 & 6.3 & 10.6 & 25.2 (3)\\
32 & 2D Torus & 3 & 4 & 0.6 & 4.9 & 9.8 & 16.5 & 39.1 \\
32 & 3D Torus & 2.5 & 6 & 0.3 & 2.7 & 5.4 & 9.2 & 21.7 \\
32 & Hypercube & 2.5 & 5 & 0.4 & 3.3 & 6.5 & 11.0 & 26.0 \\
32 & Switched & 1 & 4 & 0.2 & 1.6 & 3.3 & 5.5 & 13.0 (4) \\
64 & 2D Torus & 4 & 4 & 0.4 & 3.3 & 6.6 & 11.2 & 26.5 \\
64 & 3D Torus & 3 & 6 & 0.2 & 1.7 & 3.3 & 5.6 & 13.2 (4)\\
64 & Hypercube & 3 & 6 & 0.2 & 1.7 & 3.3 & 5.6 & 13.2 \\
64 & Switched & 1 & 4 & 0.1 & 0.8 & 1.7 & 2.8 & 6.6 (5) \\
\hline
\end{tabular}}
{\smallskip \\
Data within () reference similar entries in table~\ref{table:fftusec}}
\label{table:a2anetwork}
\vspace{-10pt}
\end{table*}

The cells in Table~\ref{table:fftusec} with parenthesized references represent particular solution choices that match well with potential communications designs, which are discussed in section~\ref{alltoall}.

\subsection{Scaling All to All}\label{alltoall}

In this section we analyze potential network topologies to evaluate points in the solution space that are compatible with distributed FFT designs.The all to all network is responsible for interchange of data among multiple processing pipelines, both when colocated on a single board and when distributed across multiple FPGAs.
 
In any parallelization into N units, $(N-1)/N$ of the data must move. Since FFT and all to all are pipelined, the overall performance will be set by the slower function.


Table~\ref{table:a2anetwork} relates numbers of FPGA modules, network topology, and the time to complete the All to All. The environment for this analysis consists of a number of FPGA nodes, each equipped with either four or six links running at 100 Gbps. Configurations marked ``Switched" use Ethernet packet framing to route messages via 100 Gbps Ethernet switches to achieve single hop connections. Each board has physical interfaces for six links, with four enabled on our test platforms.
PtoP configurations use point to point cables, other topologies require on-FPGA switches and higher hopcounts. As an example, a 4 dimensional hypercube for 16 nodes has an average hopcount of 2, because on average a destination node ID differs in only two bits from the source node ID.

The time to complete figures are given by

$$D * {\frac{{N-1}}{N}} * {\frac{H}{B * N * L}}$$

where D is the FFT data volume in bits. ${(N-1)/N}$ is the fraction of data that must be sent to a different board. H is the average hop count, B is the link bandwidth in bits per second, L is the number of links per board, and N is the number of boards that partition the problem.

The scaling behavior of this equation is such that the time to completion goes as H/N, assuming equal link loading, uniform traffic, and perfect link scheduling. All to all certainly has uniform traffic and there is a symmetry argument for uniform link loading.

For a switched network, perfect scheduling of an all to all is straightforward. In round $i$, node $n$ transmits to node $(n+i) \mod N$. Point to point networks are also easy to schedule. As for multihop scheduling, the question is still open but we are experimenting with static scheduling.

 For four boards, direct point to point cables suffice, but we must use an on-FPGA router for multihop cases. The prototype router uses a crossbar switch with buffering at each crosspoint, together with a statically compiled switch schedule. This permits the network to operate in a streaming mode without packet headers or frame boundaries. Links from application logic to the router use Intel's OpenCL Channel extension, as do the off-board links themselves.

Because the all to all communication pattern is symmetric, switch scheduling reduces to a bin packing problem of packing message fragments into open channel time slots, while managing the maximum buffer occupancy~\cite{annexstein1990unified,Subramoni2014DesigningTC}. There is no danger of livelock or deadlock and no need for traditional techniques such as virtual channels, because all messages are known at compile time. Flow control and error recovery are provided by the vendor Board Support Package (BSP). We expect to report results in future work.

\subsection{Balancing Computation vs Communications}

Table \ref{table:fftusec} and Table \ref{table:a2anetwork} identify consistent points in solution space for 8 through 128 pipelines where computation performance is a good match for communications performance. Our current configuration is (1), eight pipelines, split among four FPGAs using point to point links. Configurations (2), (3), (4), and (5) identify points in solution space for 8, 16, 32, and 64 FPGAs. These solution points permit balanced designs, in which no unit runs faster than necessary.

This analysis is done for 128x128x128 transforms, but similar choices exist for other size FFTs. Since the completion times are entirely bandwidth limited, they scale only with the total data volume and the same points in solution space apply to all sizes. However, for the smaller transforms, hardware unit latencies and communications latencies start to become important.

For $A$ atoms and 3D FFT size $N^3$, the charge spreading step operates in $O(A)$ time, the 3D FFT is $O(N^3 log N)$, and the force interpolation is again $O(A)$. By choosing an FFT volume proportional to the number of atoms (reasonably uniform density), and by using hardware to remove the $log N$ term, the entire LR pipeline becomes $O(A)$, a dramatic improvement over $O(A^2)$ pairwise methods of computing electrostatic forces.
  \section{System Implementation}
\label{systemimplementation}
\label{sec:implementation}
\subsection{FFT}

Intel provides an OpenCL computational kernel example using the feedforward parallel FFT of Garrido et al, and we took it as our starting point.\cite{intel:fft,garrido2011pipelined} (This is also used in \cite{fft3d-fpga}). This design accepts vectors bit-reversed by lane and in order by vector. 

The internal structure of the parallel FFT is shown in Figure \ref{fig:CoreFFT}. This example is 8-wide, compiled for a 64 point FFT. The design compiles a variable number of stages, corresponding to the base 2 logarithm of the transform size. As a result, the $log N$ term in FFT's $O(N log N)$ is subsumed by hardware and the unit runs in $O(N)$ time.

\begin{figure}[b]
  \vspace*{-0.2truein}
  \centering
  \includegraphics[width=\linewidth]{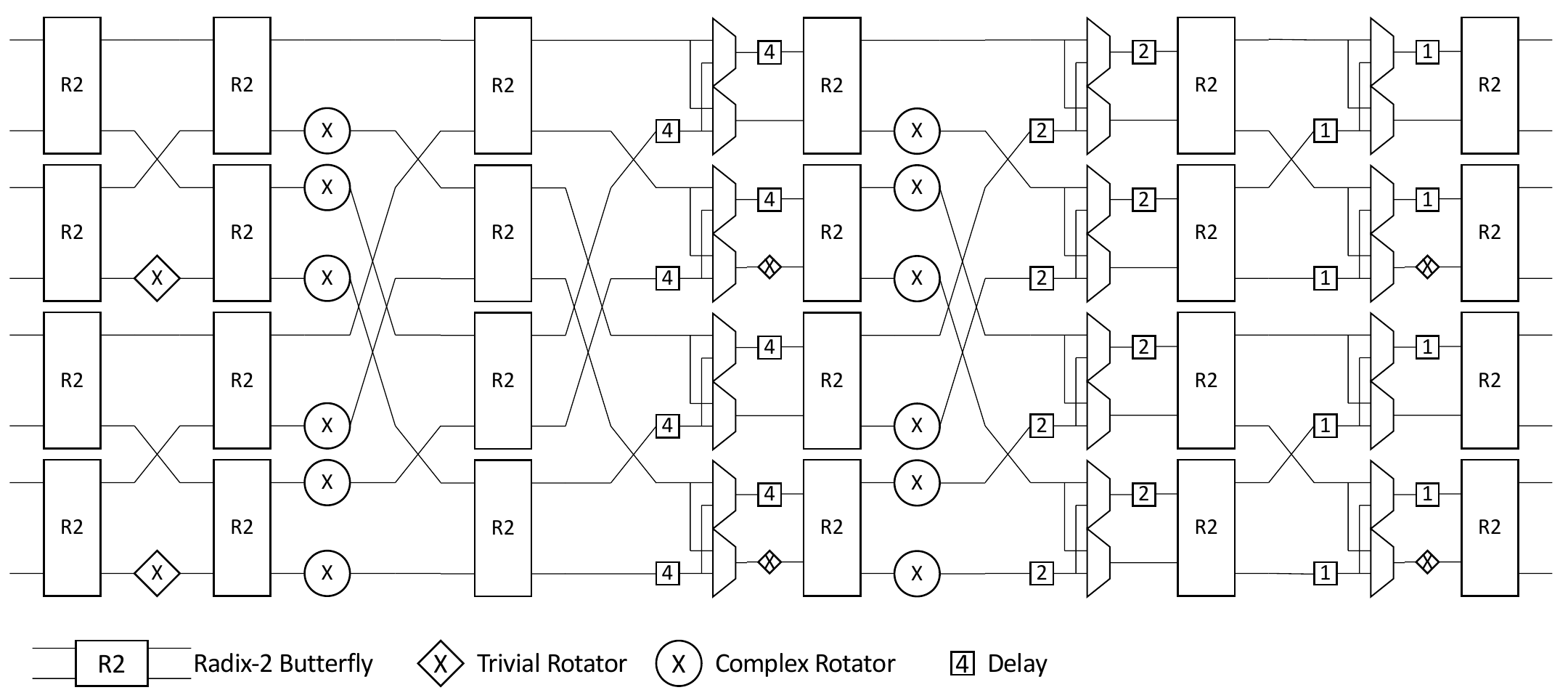}
  \vspace*{-0.2truein}
  \caption{8-Wide, single-precision complex FFT compute unit (64 point).}
  \label{fig:CoreFFT}
\end{figure}

\subsection{Bit Dimension Permutations}

In order to assemble 1D FFT units into a pipelined 3D FFT the sequencing of the data must be modified in order to deliver to and accept data from the FFT units in the correct order. This problem is referred to as bit dimension permutations~\cite{garrido2019optimum}.

 This structure is able to perform nearly arbitrary bit dimension permutations of the input sequence. A standalone python program uses a control file to generate the OpenCL source code necessary to control the hardware. 
 The limitations that exist can be removed by using 5-stage Benes networks~\cite{1675164} at the input and output but we have found the three stage networks suffice for the permutations encountered in the  3D FFT. 
 
\begin{figure}[t]
  \centering
  \includegraphics[width=\linewidth]{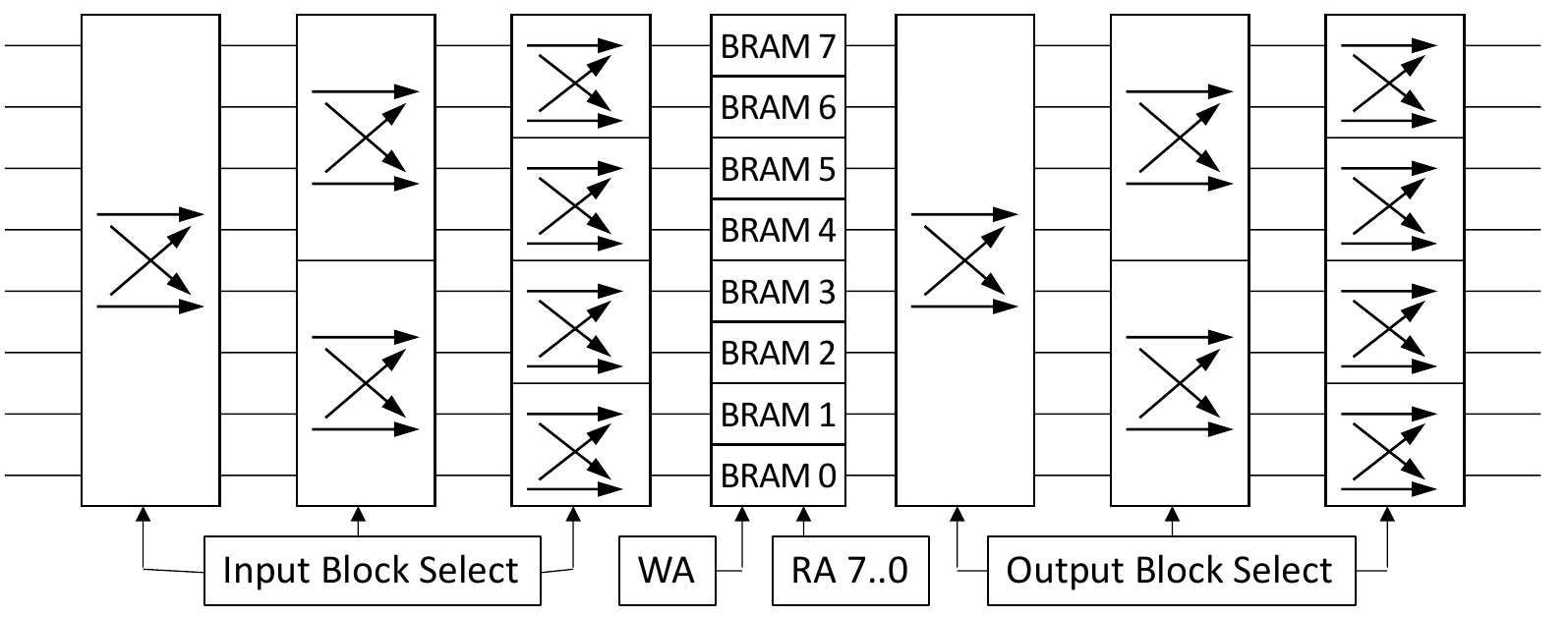}
  \caption{Transpose Unit.}
  \label{fig:transpose}
  \vspace*{0 truein}
\end{figure}

 \subsection{All to All} 
 
The All to all network is responsible for interchange of data among multiple processing pipelines, both when colocated on a single board and when distributed across multiple FPGAs.
 





Figure~\ref{fig:A2A_4_2} is the design for 8 pipelines distributed across four FPGAs. Each board has a direct connection to each other board. Some connections remain on board, but in the 4 board example they are modelled as a loopback cable that connects a board to itself. In order to use the same bit file on each of the four FPGAs, additional logic in the A2A unit routes these ``virtual cables" to the correct external port or internal loopback. This internal crossbar switch is a prototype of the future hardware supporting routed networks.

\begin{figure}[t]
  \centering
  \includegraphics[width=0.6\linewidth]{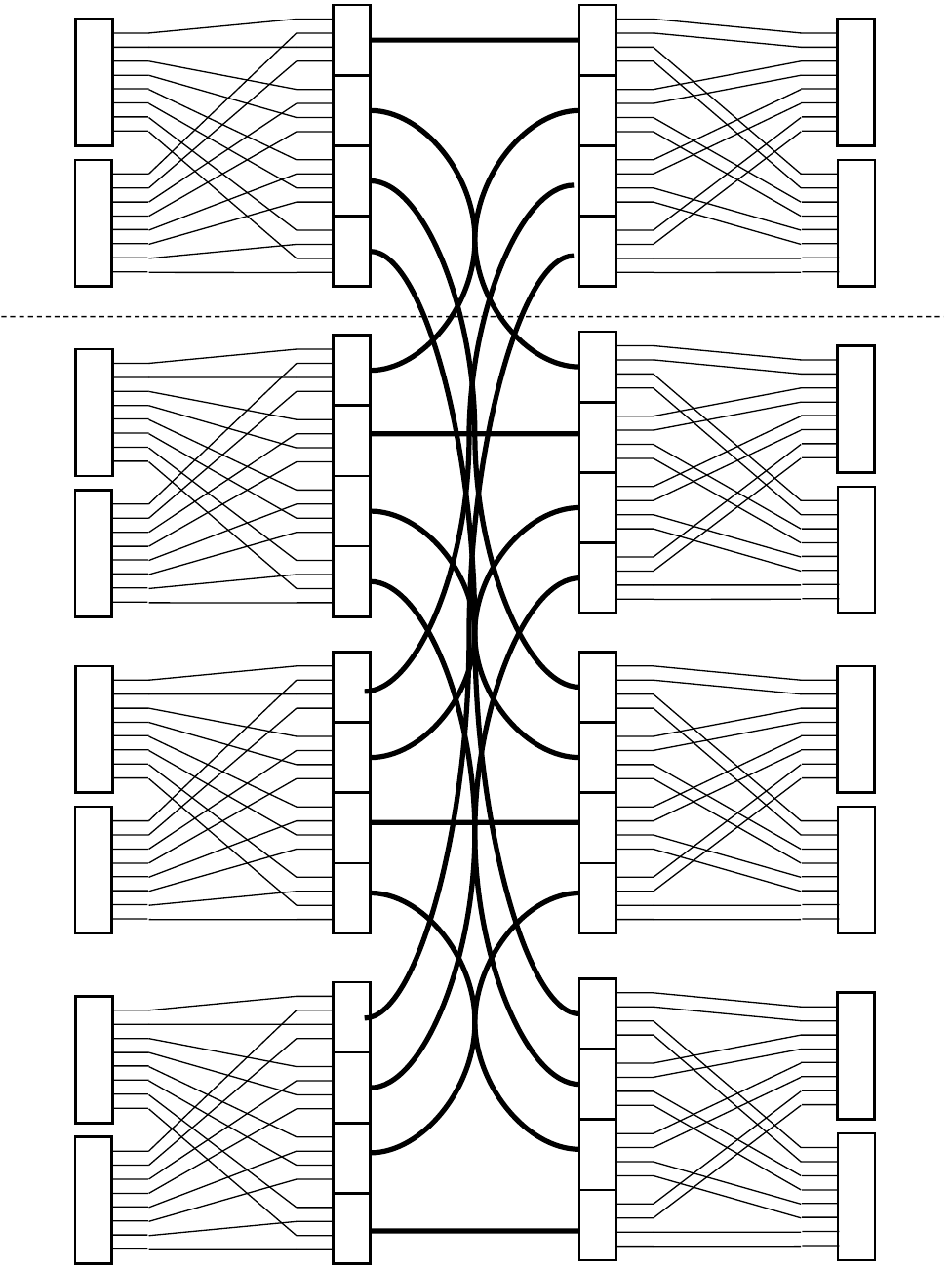}
  \caption{All to all network. 8 pipelines on four FPGAs, with cables.}
  \vspace{-10pt}
  \label{fig:A2A_4_2}
\end{figure}

\subsection{Using OpenCL for FPGA programming}
We chose OpenCL as the programming language for our implementation. This is due to several reasons: OpenCL allows more productive software development, and also allows us to be vendor-agnostic. Using OpenCL on the host side has allowed us to reuse OpenMM's framework for launching kernels on the FPGA as well. OpenCL compilation for FPGAs transforms high-level source code into a dataflow graph and instantiates the necessary hardware. We approach FPGA coding in OpenCL with a hardware engineer's perspective. It is possible to visualize the dataflow hardware you want, as in Figure~\ref{fig:CoreFFT} or Figure~\ref{fig:transpose} and then write fairly straightforward code to realize it. It can be complex to achieve the same level of control as with HDL; however, we have largely been able to overcome challenges. 

In the future, we may move some parts of our implementation into VHDL or Verilog for optimal resource utilization. OpenCL does permit linking to HDL
provided the HDL modules provide certain prescribed interfaces.

  \section{Evaluation}\label{evaluation}
\label{sec:evaluation}
We report here on two implementations. Both are in OpenCL, with no Verilog or VHDL components. The first is a FFT only design, with up to 16 processing pipelines per FPGA. The second is an implementation of Smooth Particle Mesh Ewald which implements the full long-range force pipeline. 

\subsection{Experimental Setup}
Our hardware setup comprises 8 BittWare 520N-MX boards on a single hardware node. The hardware node has 2 8-core Intel Xeon Silver CPUs as well as 768 GB of memory (used mostly for OpenCL compilation jobs). The CPUs also serve as host processor for the OpenCL programs and OpenMM. Each 520N-MX has  a single Intel Stratix 10 MX2100 FPGA~\cite{IntelStratix10} and 4 QSFP28 channels, each capable of communicating at a peak bandwidth of 100 Gb/s. Each FPGA is configured with the p520\_max\_m210h BSP to allow OpenCL as the programming model for FPGA computation as well as communication between the different boards. We are operating with a preliminary BSP which runs the links at 78 Gbps. Our application source code is compiled using Quartus release 20.3.  The server runs CentOS 7.6. We use \emph{SLURM} to manage both compilation and hardware resources. 


  The interconnect topology is shown in Figure~\ref{fig:8board}. The four boards implementing the long-range subsystem, marked LR0 to LR3 are fully connected and implement the full SPME algorithm.  Each board has a link to the corresponding short-range board.  The short-range boards are connected in a ring and are responsible for short range forces as well as motion update and force accumulation.

\begin{figure}[tb]
  \centering
  \includegraphics[width=0.8\linewidth]{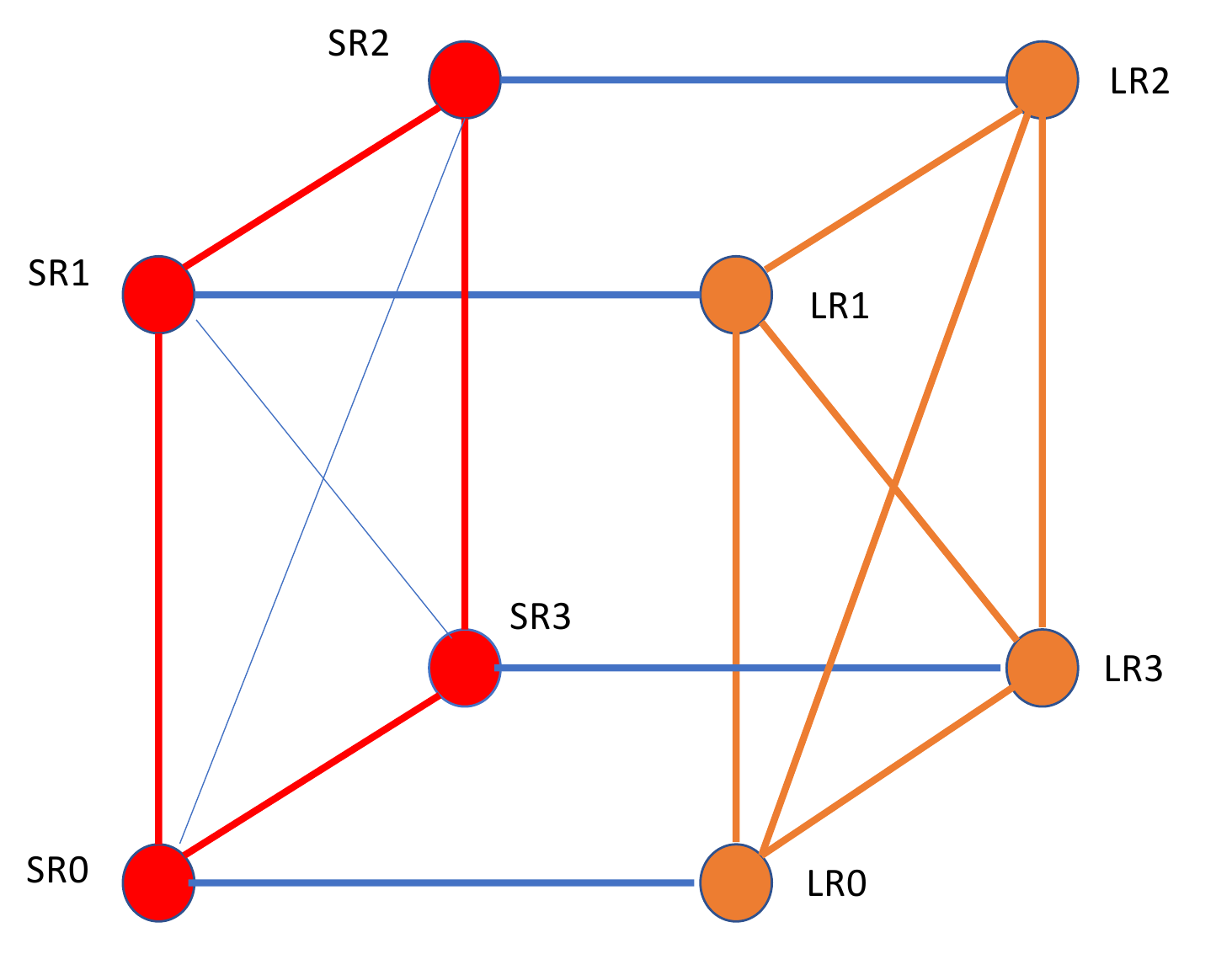}
  \vspace*{-0.2truein}
  \caption{Eight FPGA Testbed}
  \label{fig:8board}
  \vspace*{-0.1truein}
\end{figure}



\subsection{FFT Only}

We first present results of our BRAM-based FFT. In this version, the entire dataset fits into the BRAM of the FPGA. Results from this design are shown in table~\ref{table:bramfft}. We present single FPGA results for this design, with up to 16 processing pipelines, for FFT sizes $32^3$ and $64^3$. A BRAM\-only $128^3$ design will not fit on our current hardware available due to BRAM limitations.
The $32^3$ version occupies 28\% of the BRAMS and 52\% of the DSP blocks and runs in 3.87 microseconds at 266 MHz. The $64^3$ version occupies 49\% of the BRAMS and 58\% of the DSP blocks and runs in 24.5 microseconds. 

This version illustrates similar performance to Anton 1's 3.7 microseconds for $32^3$ transforms, albeit 10 years later.  We wryly note that collapsing a system of 512 ASICs into a single FPGA is fully consistent with Moore's law.


\begin{table}[t!]
\caption{BRAM-based 3D FFT}
\vspace{-6pt}
\centering
\begin{tabular}{ |c|c|c|c|c|c|c| } 
 \hline
 FFT & No. of & fMax & Time & Ideal & BRAM  & DSP  \\ 
 Size & Pipes & (MHz) & ($\mu$s) & ($\mu$s) & (\% usage) & (\% usage) \\ 
 \hline
   32x32x32 & 1 & 290 & 59 & 42 & 2 & 3 \\
   32x32x32 & 8 & 243 & 8.5 & 6.3 & 21 & 18 \\
   32x32x32 & 16 & 266 & 3.87 & 3.27 & 28 & 52 \\
  64x64x64 & 16  & 275 & 24.5 & 22.3 & 49 & 58 \\ 
 \hline
\end{tabular}
\vspace{-10pt}
\label{table:bramfft}
\end{table}

In Table~\ref{table:bramfft}, the column labelled ``Ideal" is the predicted runtime if the design were able to deliver results at exactly the compiled speed. There are two reasons for measured runtimes that are slower than ideal. First, loop dependencies may prevent the OpenCL compiler from generating a full dataflow design that can accept new operands every cycle. In OpenCL this is known as the initiation interval and the ideal value is 1. All of our designs achieve this goal. Second, unit pipeline latency and data dependency latency in the transpose unit impose delays that occur once per pass through the hardware. These effects are identifiable because they affect small transforms such as $32^3$ much more than larger ones.


\subsection{Full LR Pipeline}

The second design is the complete implementation of the long range portion of the Smooth Particle Mesh Ewald algorithm. It includes charge spreading, b-spline calculation, atom reordering for hazard suppression, forward and backwards 3D FFT, and force interpolation.

\begin{figure}[t]
  \centering
  \includegraphics[width=\linewidth]{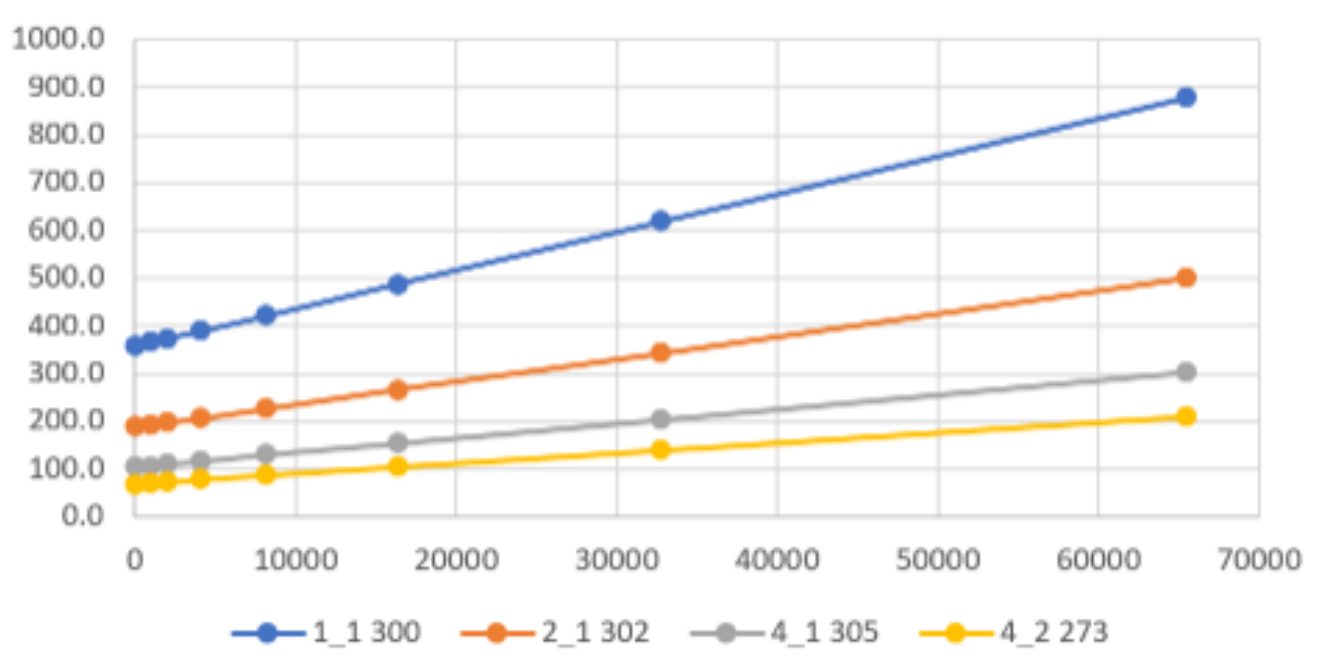}
  \vspace*{-20pt}
  \caption{Long-range pipeline performance ($\mu$s).}
  \vspace*{-4pt}
  \label{fig:LRPerf}
\end{figure}

This design is nearly all BRAM based. It uses HBM memory only for storing Green's function data. We found that any deviation from sequential memory access causes substantial degradation in memory bandwidth when using HBM. With four FPGAs, $128^3$ problems will not fit.  We can use HBM buffering in these cases, but at a performance cost.  $128^3$ will fit in BRAM for configurations of 8 FPGAs or above.

Figure~\ref{fig:LRPerf} shows the performance of the full long-range pipeline, including charge spreading, $64^3$ forward and inverse FFT, and force interpolation.  The vertical axis is in microseconds and the horizontal access reports the size of the problem in atoms.  The four lines represent 1 board 1 pipeline, 2 boards 1 pipeline each. 4 boards 1 pipeline each, and 4 boards 2 pipelines each.  The 4\_2 configuration does not scale perfectly with respect to 4\_1 because the clock speed is lower (273 MHz) and because charge-spreading performance is cable bandwidth limited when there are two processing pipelines per board.  An additional source of imperfect scaling is that as more pipelines are added, a greater proportion of atoms overlaps the boundaries between pipelines.  At 16 FPGAs, we plan to reverse this effect by doubling the number of CS and FI pipelines because at that scale each one will require fewer resources.

Figure~\ref{fig:LRPerf} can be used to read out FFT performance by considering the 0-atom case, but we measure this directly.  We have included extra hardware in the design to obtain cycle accurate timestamps for events such as last-charge-spread-atom and first-force-interpolation-atom, that bracket the FFT computation.  These results are shown in Table~\ref{table:lrfft}.
The ideal column is the minimum possible at the given fMax, with no allowance for unit latency or communications.  Current "extra" time is about 5000 clock cycles but this has not been optimized. About 40\% of the excess is due to slower than necessary transpose units. When comparing Table~\ref{table:bramfft} and  Table~\ref{table:lrfft} bear in mind that the full LR results of Figure~\ref{fig:LRPerf} and Table~\ref{table:lrfft} include two 3D FFTs and are distributed across multiple FPGAs.

\begin{table}[t!]
\caption{3D FFT Portion of LR Subsystem}
\vspace{-6pt}
\centering
\begin{tabular}{ |c|c|c|c|c| } 
 \hline
 Size & B\_P & fMax & $\mu$s & Ideal  \\ 
  \hline
  32x32x32 & 1\_1 & 313 & 67 & 39 \\
  32x32x32 & 2\_1 & 297 & 36 & 21  \\
  32x32x32 & 4\_1 & 312 & 24 & 10  \\
  32x32x32 & 4\_2 & 289 & 16 & 5.3  \\
  64x64x64 & 1\_1 & 311 & 348 & 321 \\
  64x64x64 & 2\_1 & 289 & 193 & 170 \\
  64x64x64 & 4\_1 & 311 & 99 & 79  \\
  64x64x64 & 4\_2 & 276 & 65 & 45  \\
 \hline
\end{tabular}
\vspace{-10pt}
\label{table:lrfft}
\end{table}

\subsection{Discussion}

The results shown in Table~\ref{table:bramfft} show close agreement between the ideal results and actual results, with the gap becoming smaller for larger problems. This is consistent with the effects of pipeline latency. As OpenCL compilers improve, we expect the pipeline delays will shrink. These figures are about 150-200 cycles for memory fetch, 134 for Transpose units, and 11 for the FFT.  Such improvements would be helpful for small transforms like $32^3$ but become much less important for $128^3$ since there is 64 times as much data. 

\subsection{Performance comparison with other architectures}

3D FFT, due to its wide applications in many areas has been benchmarked extensively on many architectures including CPUs, GPUs and ASICs. Many of the benchmarks focus on larger FFTs ($256^3$ and above) but there is some public information on smaller FFTs applicable to MD. Table \ref{table:comparison} compares performance of our FPGA FFT with  CPUs, GPUs and Anton. For converting timing to flops we use $15N^3lg(N)$ for complex FFT. 

For GPU measurements, we have depended both on in-house experiments as well as performance benchmarks from \cite{Sunderlanda2012AnAO,v100perf}. Our GPU code uses CUDA \emph{cuFFT} library\cite{nvidia:cufft} for computing FFT. We test this code on a single V100 GPU\cite{nvidia:v100} with CUDA 11.1 compilers and libraries. Our in-house experiments performed on V100 with NVLINK2 as well as \cite{v100perf} show that using multiple GPUs does not improve performance of sizes up to $128^3$. 

Anton 1~\cite{anton1} has details of timings for both $32^3$ and $64^3$ on 512 nodes, which we have also included in the table. 

We also include CPU-based benchmarks in this table. \cite{intelperf} shows the timings using Intel MKL and FFTW on a 56 core Intel Xeon Platinum processor. For sizes $32^3$, $64^3$ and $128^3$, performance on the processor is approximately 200, 400 and 600 GFlops respectively. We have not found many public benchmarks on performance of smaller distributed 3D FFT. We have included timings on JUGENE, a BlueGene/P architecture~\cite{10.5555/1375990.1376008}. \cite{Sunderlanda2012AnAO} shows 12 milliseconds for a problem of size $128^3$ on 512 BlueGene/P nodes. 

We have also included Novo-G's timings of distributed 3D FFT on 8 Stratix V FPGAs\cite{lawandediss} for comparison. 
Compared to other architectures in Table~\ref{table:comparison}, we outperform all architectures for $32^3$ and $64^3$ except 512 nodes of Anton 1, and V100 cuFFT for  larger sizes such as $128^3$.  

There is, as far as we know, no magic to achieving excellent 3D FFT performance. It is a game of balancing computation, memory bandwidth, and communication.  It should not be a surprise that custom ASICs can do well, nor that modern GPUs like the V100 can achieve more than a teraflop once the problem size grows large enough to sustain efficient memory access (V100 has about 50\% more flops than a Stratix 10 FPGA and almost double the memory bandwidth~\cite{nvidia:v100,IntelStratix10}). The attractive features of FPGA designs are that they can run efficiently across a range of sizes and that one can connect compute pipelines directly to communications resources, which means that one can relatively easily distribute a parallel implementation across multiple FPGAs.

\begin{table}[t]
\caption{FFT GFlops/s for multiple architectures}
\vspace{-6pt}
\centering
\begin{tabular}{ |c|c|c|c|c|c| } 
 \hline
 Size & Size & Size & Size & System & Citation \\
 $32^3$ & $64^3$ & $64^2$x128 & $128^3$ &  & \\
\hline
647 & 963 & - & - & BRAM 16 pipe & Table \ref{table:bramfft} \\
  - & - & 969 & 810 & HBM 8 pipe & Inhouse tests \\
  664 & 1774 & - & - & Anton-1 512 nodes & \cite{anton1} \\
    109 & 139 & - & 180 & Novo-G 8 FPGA & \cite{lawandediss}\\
 218 & 1358 & 1561 & 1247 & V100 cuFFT & Inhouse tests \\
  180 & 400 & 500 & 610 & 56C Xeon 8280L & MKL \cite{intelperf} \\
   - & - & - & 9 & BG/P 512 nodes& \cite{Sunderlanda2012AnAO} \\


 \hline
\end{tabular}
\vspace{-12pt}
\label{table:comparison}
\end{table}

Regarding the full implementation of PME we have more limited comparisons.  As reported above, our full LR pipeline currently completes a $64^3$ 64K Atom problem in 206 microseconds on four FPGAs. Running on a GEForce RTX 2080TI completes the equivalent phases in 523 microseconds. As discussed above, we note that the FPGA implementation scales readily with cluster size while the GPU implementation does not.

  \section{Conclusion and Future Work}
\label{sec:conclusion}
In this paper, we demonstrate that FPGAs can implement Particle Mesh Ewald in a scalable way, even for the small 3D FFTs applicable in molecular dynamics. The results show that our architecture and implementation balances computation, memory bandwidth, and communications bandwidth to produce implementations that run efficiently across multiple FPGAs. Our implementation works for a variety of molecule sizes and FFT grid sizes, and is completely written in OpenCL for portability and flexibility. Our results show that we outperform or are competitive with a wide variety of architectures including CPUs, GPUs, and ASICs. 


The goal of our work is to achieve strong scaling for FFT and the long range pipeline for molecular dynamics on multiple FPGAs. We plan to grow our FPGA cluster to 16 FPGAs, and present results on scalability of both FFT and the long-range pipeline. A larger cluster will also allow us to use BRAM only on FPGAs for $128^3$ transform as well. We also plan to explore avenues such as reducing precision for communications, exploring different layouts of the FFT dataset as well as linking VHDL/Verilog code with OpenCL. 

\bibliography{main.bib,md_refs_200628,caad_refs_201221}
  \end{document}